\documentclass[superscriptaddress, reprint, amsmath, amssymb, aps, prb, showkeys,longbibliography]{revtex4-2}
\usepackage{graphicx}
\usepackage{braket}
\usepackage{physics}
\usepackage{bm}
\usepackage{hyperref}
\hypersetup{
    colorlinks=true,
    linkcolor=blue,
    citecolor=blue,
    urlcolor=blue,
    pdfpagemode=FullScreen,
    }

\begin{document}
\title{Fractional excitations in Kitaev quasi-one-dimensional chain}
\author{Ritwika Majumder}
\email{ritwika.majumder@niser.ac.in}
\affiliation{School of Physical Sciences, National Institute of Science Education and Research, Jatni, 752050, India }
\affiliation{Homi Bhabha National Institute, Training School Complex, Anushaktinagar, Mumbai, 400094, India}

\begin{abstract}
The Kitaev honeycomb model has attracted significant interest due to its quantum spin liquid ground state, fractionalized Majorana excitations, and topological properties. Motivated by these features, we introduce a quasi–one–dimensional Kitaev–like spin chain derived from a truncated honeycomb geometry. The resulting Majorana band structure contains both dispersive and flat bands, with a gap closing at a critical anisotropy that separates trivial and chiral topological phases. Under open boundary conditions, the topological regime hosts zero–energy edge modes protected by chiral symmetry, placing the system in the BDI symmetry class with a quantized winding number. In the excited spectrum, localized plaquette modes emerge near zero energy and can be tuned by introducing domains of negative plaquettes. The dynamical spin correlation function reveals broad continua associated with fractionalized excitations, together with characteristic low–energy spectral weight from edge Majorana modes. These features distinguish the present system from conventional Heisenberg spin chains and provide experimentally relevant fingerprints of chiral topological order. Our model thus establishes a conceptual bridge between the two–dimensional Kitaev spin liquid and Kitaev’s one–dimensional quantum wire.
\end{abstract}

\maketitle
\newpage
\section{\label{sec:level1}Introduction}

\par In quantum spin systems, the investigation of bond–dependent anisotropic interactions has become one of the central themes in condensed matter physics due to their ability to host exotic quantum phases. Among these, quantum spin liquids \cite{broholm} and fractionalized excitations have been extensively studied within the framework of Anderson’s resonating valence bond theory\cite{Anderson,baskaran2017resonatingvalencebondtheory}. A major breakthrough was achieved through Kitaev's work on an anisotropic spin-interacting model \cite{KITAEV20062}, which admits a ground state with a highly entangled magnetically disordered spin state without any symmetry breaking, referred to as a quantum spin liquid. In the presence of an extensive number of conserved quantities, the mapping of Majorana fermions and gauge fields through enlarging the Hilbert space makes this model integrable. The model maps exactly onto itinerant Majorana fermions moving in a static $\mathbb{Z}_2$ gauge background \cite{RevModPhys.80.1083}. This motivates the search for additional interacting models that remain exactly solvable \cite{PhysRevB.84.155121,PhysRevLett.98.247201,PhysRevLett.127.127201}. Unlike the Landau theory, these highly entangled quantum phases are not characterized by magnetization or local structure factors. Instead, they are often characterized by topological order parameters and signature fractionalized excitations \cite{yoshida2013exotic,wen1991mean,PhysRevB.65.165113,PhysRevB.94.195130}. Depending on the coupling strengths, the Kitaev model shows gapped and gapless spin liquid phases with nontrivial topological order \cite{Savary_2017}. Thus, the model establishes an important bridge between strongly correlated spin systems and topological band theory through the emergence of Majorana quasiparticles and protected edge states. In the presence of time–reversal symmetry–breaking perturbations, the gapped phase supports non-Abelian anyonic excitations, generating immense interest for applications in fault-tolerant topological quantum computation \cite{RevModPhys.80.1083, KITAEV20032}.   

The search for physically realizable Kitaev models has driven extensive theoretical and experimental activity over the last decade. Spin–orbit coupled Mott insulators containing heavy transition metals such as iridium and ruthenium have been proposed as candidate materials where bond–dependent exchange interactions emerge naturally through Jackeli–Khaliullin mechanisms \cite{PhysRevLett.102.017205}. In particular, honeycomb compounds such as $\alpha$-RuCl$_3$, Na$_2$IrO$_3$, and Li$_2$IrO$_3$ have been widely investigated as proximate Kitaev quantum spin liquids \cite{trebst2017kitaevmaterials,Winter_2017}. Experimental probes, including neutron scattering, Raman spectroscopy, and thermal transport, have revealed signatures of fractionalized excitations and broad scattering continua that resemble predictions from the Kitaev model \cite{Banerjee2016,3m4m-3v59,Kasahara2017,Kasahara2017,Kasahara2018}. Although the presence of additional Heisenberg and
off–diagonal interactions may often destabilize the exact spin liquid phase in real materials, these systems continue to provide compelling evidence for emergent Majorana physics \cite{Zhang2025}.

\begin{figure*}[t]
    \centering
    \includegraphics[width=\linewidth]{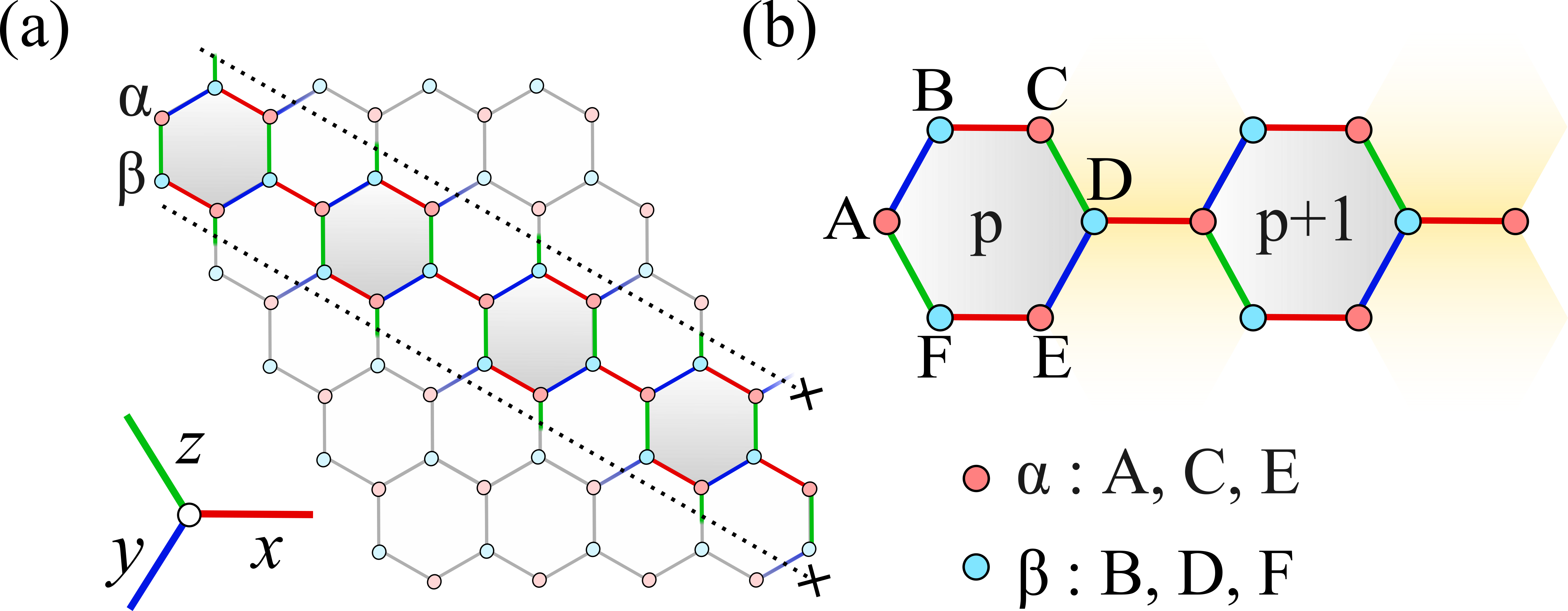}
    \caption{(a) Schematic representation of the quasi–one–dimensional hexagonal chain layer embedded within the honeycomb geometry. Each hexagonal plaquette (grey) consists of 3 pairs of two sub-lattice classes connected by three different nearest–neighbor bonds. Red, blue, and green bonds denote anisotropic spin couplings $S^{x}S^{x}$, $S^{y}S^{y}$, and $S^{z}S^{z}$, respectively. (b) Isolated chain segment comprising plaquettes $p$ and $p+1$, where each plaquette (unit cell) contains six sub-lattice sites grouped into two distinct classes as indicated.}
    \label{fig:0}
\end{figure*}

Another defining feature of Kitaev systems is the emergence of fractionalized dynamical
excitations\cite{PhysRevLett.124.017201,hermanns2018physics,wang2025fractionalization,theveniaut2017bound}. Unlike conventional magnets, where spin flips generate sharp magnon modes, the dynamical spin structure factor in Kitaev models exhibits broad continua arising from the fractionalization of spins into Majorana fermions and flux excitations, distinct from conventional spinon continua \cite{theveniaut2017bound,PhysRevLett.113.187201,PhysRevLett.112.207203}. Such continua have become a hallmark signature of quantum spin liquids and are now routinely used as diagnostics in both theoretical and experimental studies. Understanding how these continua evolve across topological phase transitions and in geometries in reduced dimensions remains
an active problem. Depending on the underlying symmetries, Kitaev systems may belong to Altland–Zirnbauer symmetry classes supporting topologically protected edge excitations \cite{PhysRevB.55.1142,PhysRevResearch.3.013052,PhysRevLett.42.1698,Kitaev_2001}. In one dimension, chiral symmetry plays a particularly important role, as exemplified by the Su–Schrieffer–Heeger (SSH) chain and Majorana superconducting wires \cite{Ryu_2010}. However, a simple Kitaev chain with $x$- and $y$-bond interactions \cite{PhysRevB.107.104414} lacks deconfined $\mathbb{Z}_2$ gauge fields and nontrivial anyonic braiding statistics \cite{GREITER20141026}. These developments have motivated the exploration of decorated spin models whose Majorana representations naturally realize one–dimensional topological phases.

Motivated by these developments, several recent works have focused on quasi-one–dimensional systems that retain essential features of Kitaev physics while allowing new forms of topology
and localization \cite{PhysRevB.82.174409,Agrapidis2018,pujari2025theoremextensivegroundstate,yamazaki2026majoranaassistednonlocalspincorrelation,Catuneanu2018,PhysRevLett.99.247203}. Ladder geometries, decorated chains, and fragmented honeycomb structures have been shown to host flat bands, localized Majorana modes, and symmetry–protected topological phases \cite{PhysRevB.109.235412,huda2020designer,martinez2024topological,huang2020quasi}. In many such systems, the reduced connectivity modifies the gauge structure and can produce unconventional degeneracies absent in the original honeycomb lattice \cite{Zhang2025,mandal2020introductionkitaevmodeli}. Furthermore, the interplay between lattice geometry and bond anisotropy often generates topological phase transitions that are more transparent than in two dimensions. 

Inspired by the structure of copper pyrazine dinitrate (CuPzN) \cite{jones2001vibrational,lancaster2006magnetic}, in the present work, we introduce and investigate a quasi–one–dimensional Kitaev–like chain obtained by systematically truncating bonds of the original honeycomb lattice \cite{KITAEV20062}. We show that the resultant geometry forms an array of coupled hexagonal plaquettes connected through anisotropic inter-plaquette links. Despite the dimensional reduction, the model preserves the essential ingredients of Kitaev physics, viz., bond-dependent interactions, local conserved plaquette operators, emergent $\mathbb{Z}_2$ gauge fields, Hilbert-space fragmentation, and an exact Majorana-fermion representation. We discuss the band dispersion, band topology, and protected edge modes in open boundary conditions (OBC). We find that the reduced connectivity leads to features, including non-dynamic Majorana modes, flat bands, and a chiral topological phase transition controlled by exchange anisotropy \cite{Zhang2025,zhuang2025gaplessorderedphasesspin12}. In the uniform positive plaquette ground state, if we grow negative plaquettes in a systematic fashion, zero modes emerge with states localized at in plaquettes at the domain boundaries. To further characterize the phases, we investigate the dynamical spin response using density matrix renormalization group (DMRG) calculations \cite{PhysRevB.107.104414}. The dynamical structure factor reveals broad continua across both trivial and topological regimes, confirming the persistence of fractionalized excitations. In the topological phase, additional low-energy spectral weight emerges due to edge Majorana modes, providing experimentally accessible signatures of the phase transition. 

This paper is organized as follows. In Sec. \ref{sec:level2}, we describe the quasi-one-dimensional Kitaev model and its various features like the Hamiltonian, its Majorana construction, and the Hilbert space. In Sec. \ref{sec:level3}, we discuss its edge modes, excited plaquette modes, fractionalized excitations, and the corresponding topological phases. We conclude in Sec. \ref{sec:level4} by discussing how the described features in previous sections distinguish the present system from conventional Heisenberg spin chains and suggest possible routes toward observing topological Majorana physics in quasi–one–dimensional magnetic materials.

\section{\label{sec:level2}Quasi-one-dimensional Kitaev Model}

\subsection{Model Hamiltonian}
\par The original Kitaev model is characterized by anisotropic nearest‑neighbor spin–spin interactions that depend explicitly on the orientation of the bonds \cite{KITAEV20062}. Each lattice site is connected to three nearest neighbors through distinct bond directions, with each bond selectively coupling a specific spin component ($S^{x}$, $S^{y}$, or $S^{z}$) between adjacent spins, as illustrated in Fig.~\ref{fig:0}(a). By cutting the $y$ and $z$ bonds along the dotted line, the lattice reduces to the isolated chain shown in Fig.~\ref{fig:0}(b). In this representation, red bonds denote couplings between $x$‑components of neighboring spins, while blue and green bonds correspond to $y$‑ and $z$‑component couplings, respectively. The resulting geometry forms an array of hexagonal plaquettes connected by inter‑plaquette bonds, and the interaction Hamiltonian can be expressed as

\begin{widetext}
    \begin{equation}
    H=\sum_{p\in hexagon}[J_yS_{pA}^yS_{pB}^y+J_xS_{pB}^xS_{pC}^x+J_zS_{pC}^zS_{pD}^z+J_yS_{pD}^yS_{pE}^y+J_xS_{pE}^xS_{pF}^x+J_zS_{pF}^zS_{pA}^z+J_xS_{pD}^xS_{p+1A}^x]
    \label{1}
\end{equation}
\end{widetext}

where $S^{\alpha}=\left(\frac{\hbar}{2}\sigma^{\alpha}\right)$'s are spin-$\tfrac{1}{2}$ operators proportional to Pauli matrices, $i$ is the plaquette index and \{A,B,...,F\} are the six sub-latices of corresponding plaquettes. Unlike the exact 1D chain as in \cite{PhysRevB.107.104414,PhysRevB.105.085106,6ytv-7d16}, in this construction, we truncate the bonds such that the plaquette structures stay and the inter-plaquette spin interactions are between the $x$-components of spins. Choosing the $z$-component as the inter-plaquette link is just a 120$^{\circ}$ space rotation while similarly truncating the other two bonds. Each plaquette has two inter-plaquette and 6 intra-plaquette couplings as shown in Fig.~\ref{fig:0}(b). From the next section onward $\hbar/2$ will be scaled to 1.

\subsection{Conserved quantity}
\par The original model has an extensive number of hexagonal plaquettes as conserved quantities \cite{mandal2020introductionkitaevmodeli}. In the new lattice discussed here, we define the plaquette operators as $W_p=\prod_{j\in p}\sigma_{pj}^{\alpha_j}$ where $\alpha_j$ are components of the $j$-th site, that are not part of the $p^{th}$ plaquette. Unlike the honeycomb structure, here all plaquettes are independent, and the larger composite loops are absent. Hence, the phase or flux of each plaquette is confined to itself. From the figure, the plaquette operator will be 
\begin{equation}
\begin{aligned}
    &W_p=\sigma_{pA}^x\sigma_{pB}^z\sigma_{pC}^y
    \sigma_{pD}^x\sigma_{pE}^z\sigma_{pF}^y\\
    &[W_p,H]=0, \qquad [W_p,W_{p'}]=0, \quad  \forall p,p'
\end{aligned}
\label{2}
\end{equation}
Hence, in a chain with $6N$ number of sites, we have $N$ plaquettes, and hence, $N$ conserved quantities. Therefore, the Hilbert space forms $2^N$ diagonal sectors for every set of $\{W_p\}$. 

\subsection{Majorana construction}
To solve the Hamiltonian, we use Kitaev's method of mapping the Pauli matrices into Majorana fermions as in \cite{KITAEV20062}. So each spin operator is defined as 
\begin{equation}
    \sigma^{\alpha}_{j}=ic_j^{\alpha}c_j
    \label{3}
\end{equation}
So, the spins at each sub-lattice site is mapped to Majorana fermions of four different flavors $c_j^x$, $c_j^y$, $c_j^z$, and $c_j$. The operators are self-conjugate and anti-commute with all other flavors. Therefore, a two-dimensional local Hilbert space gets mapped to a four-dimensional enlarged local Hilbert space. Correspondingly, the Hamiltonian in \eqref{1} gets mapped to a new Hamiltonian $\tilde{H}$ with a doubled local Hilbert space. In the physical subspace of the extended Hilbert space the spin algebra will be satisfied. Hence, $\sigma^x_j\sigma^y_j\sigma^z_j=i$ will hold in that space. Under the Majorana mapping, $\sigma^x_j\sigma^y_j\sigma^z_j\rightarrow ic_j^xc_j^yc_j^zc$ can have values $\pm i$. We can define an on-site projector as
\begin{equation}
    \begin{aligned}
        D_j=c_j^xc_j^yc_j^zc_j, \quad & P_j=\frac{1+D_j}{2}
    \end{aligned}
    \label{4}
\end{equation}
where $\{I,D_j\}$ form a $Z_2$ gauge group and $P_j$ is 1 in the physical subspace and 0 otherwise. Hence, this projects the operators satisfying the spin-$\tfrac{1}{2}$ algebra. Since $[\tilde{H},P_j]=0\quad \forall ~j$, the eigenstates can first be obtained for $\tilde{H}$ and then be projected to the actual physical eigenspace.

\begin{figure}
    \centering
    \includegraphics[width=\linewidth]{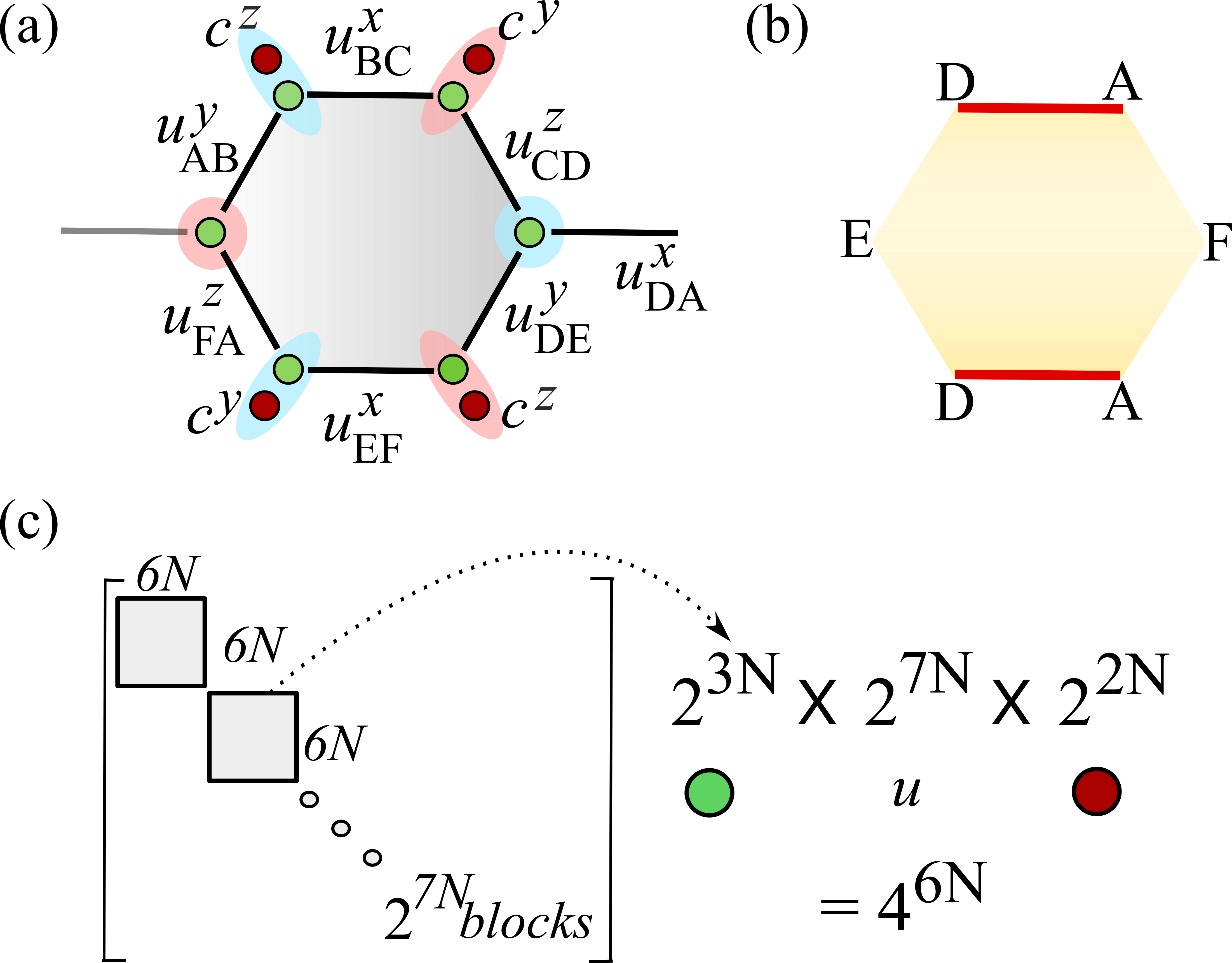}
    \caption{(a) Elementary hexagon with bond variables $u_{\alpha\beta}$ indicated on each link. The itinerant Majorana fermions are shown as green circles, while the non–dynamic Majoranas are represented by brown circles. On each bond, a gauge variable $u_{\alpha\beta}$ is defined. (b) Illustration of incomplete plaquettes inverted relative to the hexagon, where the DA bond appears twice within the plaquette structure. (c) For a chain of $N$ unit cells, the Hamiltonian dimension is $6N\times 6N$ for a fixed configuration of $u_{\alpha\beta}$. Each unit cell contains seven bonds, leading to $2^{7N}$ distinct Hamiltonian blocks. In addition, the $2^{2N}$ degeneracy arising from the non–dynamic Majoranas (colored brown) further fragments the Hilbert space.}

    \label{fig:1}
\end{figure}

We can represent each term of the Hamiltonian in the extended fermionic space, say, $\sigma_{iA}^y\sigma_{iB}^y\rightarrow(ic_{iA}^xc_{iB}^x)(ic_{iA}c_{iB})$ similarly for all terms in eq.\eqref{1}. In this representation, at each site, the spin-$\tfrac{1}{2}$ is represented by four different flavors of Majorana fermions, i.e., $c^x,c^y,c^z,c$. While all four Majorana flavors in a honeycomb stacking participate in dynamics, in our case, we find that only the sites $A$ and $D$ associated with the inter-plaquette links display this behavior. The other four sites ($B,C,E,$ and $F$) have one extra flavor of Majorana per site, the brown ones in Fig. \ref{fig:1} (a), which do not appear in the Hamiltonian in eq.\eqref{5}. This feature also appears under OBC in two dimensions \cite{Zhang2025}. So the mapped Hamiltonian becomes
\begin{equation}
    \tilde{H}=\sum_{p\in plaquette}\sum_{\langle j,k\rangle_{\alpha}\in p}(J_{\alpha}u_{jk}^{\alpha})ic_{pj}c_{pk}+(J_xu_{DA}^{x})ic_{pD}c_{p+1A}
    \label{5}
\end{equation}

where $u_{jk}^{\alpha}=ic_{pj}^{\alpha}c_{pk}^{\alpha}$ and $u_{DA}^{x}=ic_{pD}^{x}c_{p+1A}^{x}$ are defined on intra-plaquette $\alpha$-bond and inter plaquette $x$-bond respectively. From the algebra, we see that all the $u_{jk}^{\alpha}$ are constants of motion in the mapped Hamiltonian. Following the Majorana fermion statistics the $u_{jk}^{\alpha}$'s satisfy, (a) $u_{jk}^{\alpha}=-u_{kj}^{\alpha}$, (b) $(u_{jk}^{\alpha})^2=1$ and (c) commutes with all other $u_{jk}^{\alpha}$. Therefore, the $u_{jk}^{\alpha}$'s can have values $\pm 1$ in their eigenspace. Hence, they now become the $\mathbb{Z}_2$ gauge field associated with each bond, mentioned in Fig.\ref{fig:1}(a). Consequently, on each site, there is only one Majorana flavor that is itinerant under the action of the Hamiltonian.

Hence, we can cast the Hamiltonian $(\tilde{H})$ in the eigen-basis of $u_{jk}^{\alpha}$. As shown in eq.\eqref{5}, it becomes a tight-binding Hamiltonian with a $Z_2$ field associated with each hopping term, making it integrable. To find the ground state, we must fix the gauge configuration on each bond, thereby fixing the plaquette values. From Lieb's theorem, we know that in a time-reversal-symmetric two-dimensional lattice with coordination number 3, the phase acquired around a loop is zero. Hence, the plaquette values $W_p=e^{i\phi_p}=1$ for all $p$. Following the theorem, $W_p=e^{i\phi_p}=1$ gives us a choice to make all $u_{jk}^{\alpha}=1$ for bonds within the hexagon plaquette. The gauge fields for intra-plaquette bonds are yet to be assigned.

\subsection{Hilbert space}
\par For a chain of $N$ plaquettes, the lattice consists of $6N$ sites, and the original spin-$\tfrac{1}{2}$ degrees of freedom provide a local Hilbert space of dimension $2$, yielding a total Hilbert space of dimension $2^{6N}$. Upon Majoranization, the dimension of the local Hilbert space doubles to $4$, but this enlarged space ($4^{6N}$) is fragmented into blocks due to conserved quantities. Specifically, the $\mathbb{Z}_2$ gauge variable on each bond admits two choices, resulting in $2^{7N}$ distinct gauge configurations because there are 7 independent fields per unit cell. For each fixed set $\{u\}$, the Hamiltonian is quadratic in Majorana space and spans a $6N$-dimensional space, producing $6N$ eigenvalues, of which $3N$ Majorana modes are filled. For a fixed gauge sector, filling the negative-energy Majorana states yields $2^{3N}$ many-body configurations. In addition, sites $B, C, E,$ and $F$ host spin components that do not participate in the Hamiltonian, giving rise to four non-dynamic Majorana quasiparticles, two for each flavor per plaquette. These four Majorana quasiparticles, grouped into two complex fermions, contribute an additional degeneracy since each fermion can be either occupied or empty. Consequently, there is a $2^{2N}$ degeneracy arising from the non-dynamic Majoranas. Altogether, the Hilbert space is fragmented into sectors characterized by gauge configurations, dynamical Majorana fillings, and residual degeneracies from non-dynamic modes, described in Fig.\ref{fig:1}(c).
\begin{figure}
    \centering
    \includegraphics[width=\linewidth]{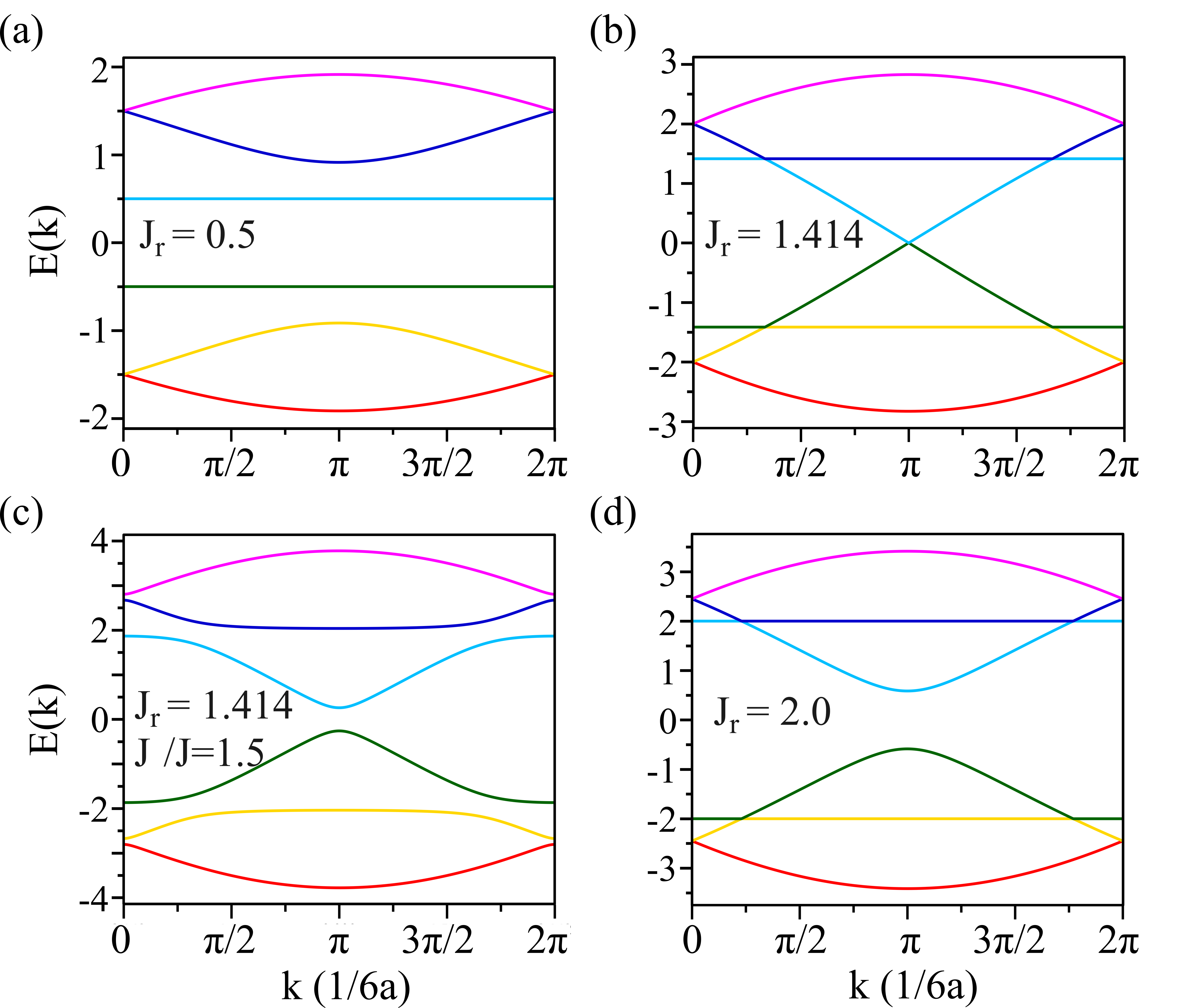}
    \caption{Majorana band structure for four representative values of $J_{x}$ with $J_{y}=J_{z}=J$. (a) $J_r=J_{x}/J=0.5$: the flat bands are separated and lie closest to the chemical potential. (b) $J_{r}=\sqrt{2}$: the band gap closes, indicating the possibility of a topological transition. At that gapless point, if we change $J_y$ to $1.5J$, a gap opens at all the band crossing points, as drawn in (c). As a result, the system becomes gapped again. (d) $J_{r}=2$: band dispersion in the gapped phase.}
    \label{fig:2}
\end{figure}
\section{\label{sec:level3}Results and Discussions}
\subsection{Ground state}
In the Majorana representation, the ground state corresponds to a chemical potential fixed at zero. Respecting the unit cell structure, two inequivalent choices remain for the $\mathbb{Z}_2$ field on the plaquette link, namely $\pm 1$ on every bond. The resulting band structure, shown in Fig.~\ref{fig:2}, consists of four dispersive bands and two flat bands. For relative value $J_y=J_z=1$, and $J_x=J_r$ and setting the all $\{u\}=1$, the 6 bands in momentum space are $\pm J_r$, $\pm\sqrt{e^{2ik}(2+J_r^2)\pm J_r\sqrt{2e^{3ik}(1+e^{ik})^2}}$.

The spectrum is gapless for $J_x/J =J_r= \sqrt{2}J$ with $J_y = J_z = J$, drawn in  Fig.~\ref{fig:2}(b), while on either side of this parameter regime a finite gap opens, as in Fig.\ref{fig:2} (a) and (d). At the gapless point, the dispersion is linear around the gap-closing point. For the choice of the inter-plaquette $\mathbb{Z}_2$ field value $+1$, the gapless point appears at momentum at the Brillouin zone edge ($\pi$), whereas for $-1$ it shifts to zero momentum. Due to this gauge flip, the whole spectrum translates by half of the Brillouin zone. The flat bands occur at energies $\pm J_r$, so for sufficiently small anisotropy ($J_x \leq J/\sqrt{2}$) the flat bands become the closest bands to the chemical potential as in Fig.~\ref{fig:2}(a). Since $J_y$ and $J_z$ correspond to equivalent bonds, they produce identical effects on the band structure. Staying at the gapless value of $J_r$, if $J_y$ is deviated from $J$, the spectrum opens up gaps at all the band crossing points. As described in Fig.~\ref{fig:2}(c), the system becomes gapped. This also makes all the bands dispersive. If we increase the $J_y$ anisotropy for a particular $J_r$ value, we find another linear gap-closing point.

The ground state remains invariant under random choices of $u_{jk}^{\alpha}$ on the inter-plaquette links. This can be verified through a real-space diagonalization with random choices of $u_{jk}^{\alpha}$ yielding the same eigenvalues. This feature can be understood by noting that each such bond belongs simultaneously to two yellow plaquettes (top and bottom), which are truncated in the construction, as illustrated in Fig.~\ref{fig:1}(b). These plaquettes contain two $x$-links that will coincide with the same bond if the truncation was weak, so any $\mathbb{Z}_2$ gauge value squares to unity and contributes zero phase to the periodic plaquette flux. Consequently, the $\mathbb{Z}_2$ gauge field on inter-plaquette links does not affect the physical spectrum and is therefore irrelevant for the dynamics of the system. 

\begin{figure}
    \centering
    \includegraphics[width=\linewidth]{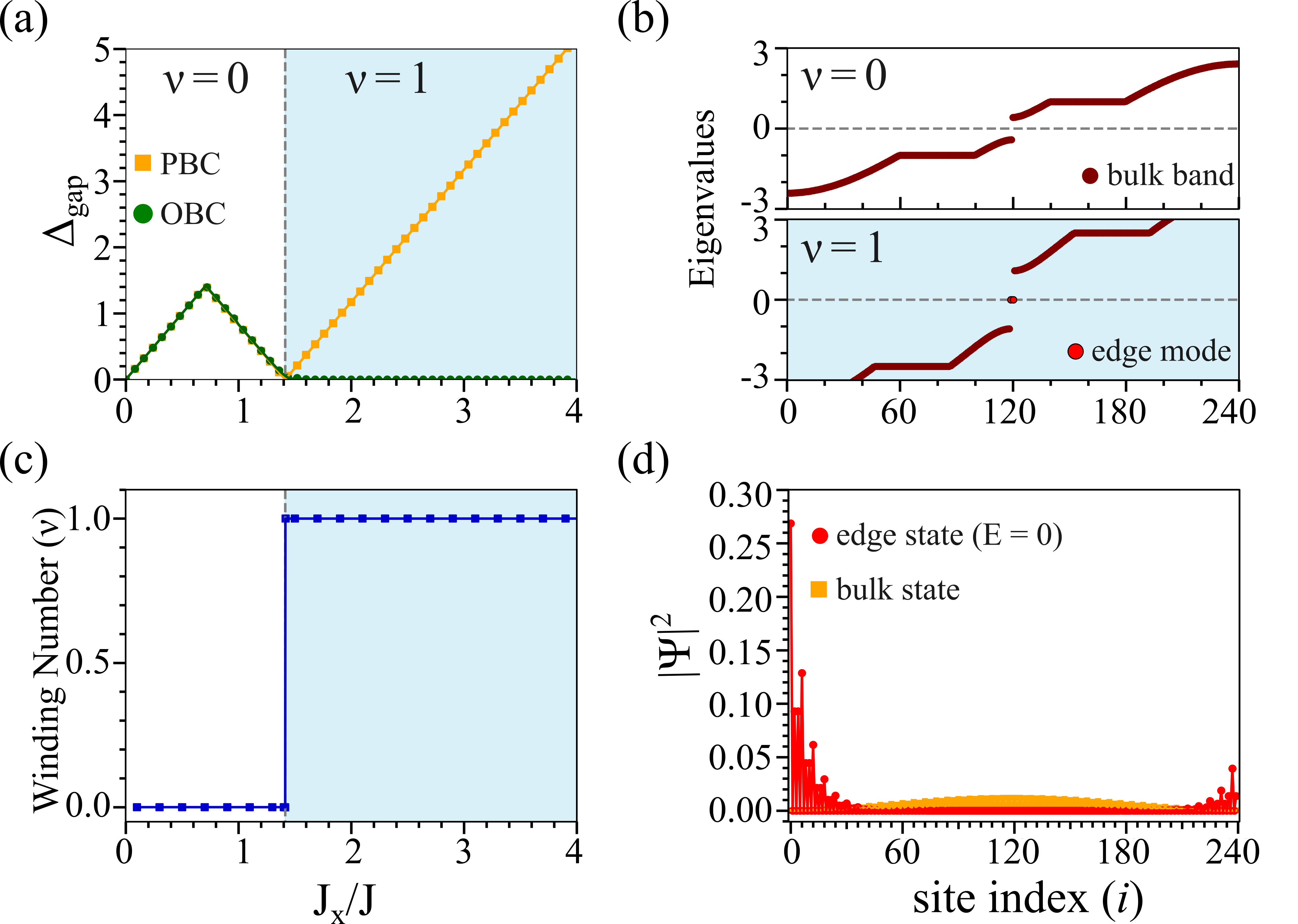}
    \caption{Topological characterization of the quasi–one–dimensional chain. (a) Band gaps at the chemical potential are shown for periodic boundary conditions (yellow) and open boundary conditions (green). (b) Eigenvalue spectra under OBC: the upper panel corresponds to the topologically trivial regime, while the lower panel illustrates the chiral topological phase with edge states pinned at the chemical potential. (c) Winding number $(\nu)$ plotted as a function of $J_{x}$, serving as the topological invariant and confirming the phase transition. (d) Probability distribution of the eigenstates closest to the chemical potential for the same parameter choice as in (b), highlighting edge localization. Calculations were performed on a 240–site chain under OBC and on 240 unit cells for PBC.}
    \label{fig:3}
\end{figure}

\subsection{Edge Modes and Topological Phase Transition}

The analysis of the single–particle spectrum revealed a nontrivial gapless point at $J_{r}=\sqrt{2}$. To probe the topological character of this transition, we examined the spectrum under open boundary conditions (OBC) for chains with commensurate sub-lattice sites. Numerical diagonalization shows that for $J_{r}>\sqrt{2}$, the bulk gap closes and two isolated zero–energy states emerge, as shown in Fig.~\ref{fig:3}(a-b). Their corresponding eigenfunctions are exponentially localized at the chain ends, confirming the presence of edge modes and signaling a topological phase transition, as shown in Fig.~\ref{fig:3}(d). 

From symmetry considerations, this quasi–one-dimensional chain belongs to the $C_{2v}$ point group. The unit cell contains two inequivalent sub-lattice sets:

\begin{equation}
    \alpha=\{A,C,E\}, \qquad \beta=\{B,D,F\}.
    \label{6}
\end{equation}

Under inversion about the center of the $x$–bond, the mapping $B,D,F \mapsto E,A,C$ holds, thereby protecting inversion symmetry. This symmetry enforces a quantized Zak phase, consistent with the observed boundary states.

The connectivity of the lattice is such that the $\alpha$–sites couple only to $\beta$–sites, with no intra–sub-lattice hopping. The Bloch Hamiltonian, therefore, takes the chiral block form

\begin{equation}
H(k)=
\begin{bmatrix}
 \mathbf{0} & \mathbf{f}(k) \\
 \mathbf{f}^{\dagger}(k) & \mathbf{0} 
\end{bmatrix},
\qquad 
\mathbf{f}(k)=
\begin{bmatrix}
 1 & J_{r}e^{ik} & 1 \\
 J_{r} & 1 & 0 \\
 0 & 1 & J_{r} 
\end{bmatrix}.
\label{7}
\end{equation}

This structure explicitly exhibits chiral symmetry. To quantify the topological phase, we compute the winding number
\begin{equation}
    \nu=\frac{1}{2\pi i}\int_{-\pi}^{\pi} dk \, \partial_{k}\ln\det[\mathbf{f}(k)],
    \label{8}
\end{equation}
which changes discontinuously across $J_{r}=\sqrt{2}$, consistent with the edge–state analysis as shown in Fig.\ref{fig:3}(c). According to Fig.\ref{fig:3}(d), in the trivial phase ($\nu=0$), no boundary modes are present, whereas in the topological phase ($\nu=1$), a pair of zero–energy states appear localized at the chain ends. These edge modes are protected by chiral symmetry, ensuring their robustness against local perturbations that respect the symmetry.

Hence, the chain is classified within the BDI symmetry class \cite{PhysRevB.55.1142,PhysRevResearch.3.013052}, analogous to the SSH model and the 2D Kitaev model. The coexistence of inversion symmetry, chiral block structure, and quantized winding number supports the occurrence of a chiral topological phase transition in this system.

\begin{figure}
    \centering
    \includegraphics[width=\linewidth]{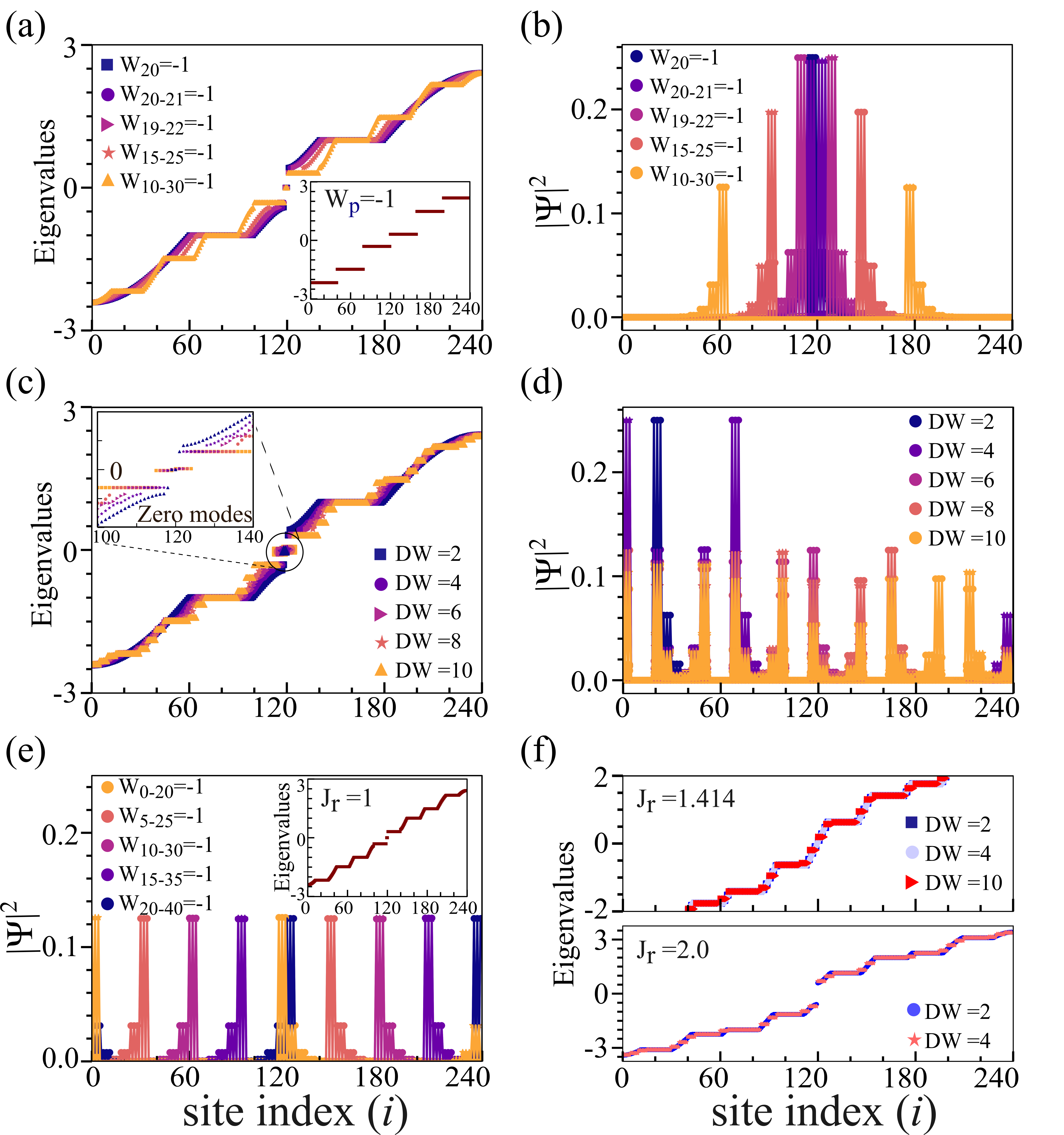}
    \caption{ $W_{p_1-p_2}=-1$ represents a negative plaquette domain from $p_1$-th to $p_2$-th plaquette on a positive plaquette background. (a) The evolution of the spectrum with 2 zero-energy states as we increase the number of negative plaquettes in a single domain. (b) The corresponding probability of the zero modes. (c) and (d) describe the evolution of the spectrum and the probability of zero modes. respectively, as we increase the number of domains of negative plaquettes in a lattice. The corresponding eigenvalues in (a) and (c) show isolated modes for $J_r=1$. (e) The probability distribution of zero modes (shown in the inset) with the translation of plaquette modes with $p_1$ and $p_2$. (f) The evolution of eigenvalues shown for different numbers of domain wall (DW), for $J_{r}=\sqrt{2}$ (upper panel) and $J_{r}=2$ (lower panel). }
    \label{fig:mode}
\end{figure}

\subsection{Plaquette Excitations and Localized Modes}

We now turn to the eigen–spectrum in the presence of plaquette excitations. In the isotropic limit, the ground state with all plaquettes fixed to $+1$ yields the band structure shown in Fig.~\ref{fig:2}, consisting of two flat and four dispersive bands. By contrast, in the excited sector where all plaquettes are set to $-1$ [Fig.~\ref{fig:mode}(a), inset], the six Majorana bands collapse into six flat bands, none of which reside at $\pm J_{x}$.

In an excited configuration where an $N$‑plaquette chain hosts $N/2$ consecutive negative plaquettes, the spectrum evolves into four dispersive bands accompanied by eight flat bands, reflecting the coexistence of positive and negative domains. Within the topologically trivial regime ($J_{r}<\sqrt{2}$), two zero‑energy modes emerge whose eigenfunctions are strongly localized at the boundary plaquettes on either side of the domain. Translating the negative domain across the lattice shifts the localization of these modes accordingly, confirming their character as domain-wall-bound plaquette states [Fig.~\ref{fig:mode}(e)].

When a single plaquette is flipped to $-1$, similar zero modes appear isolated from the rest of the spectrum under periodic boundary conditions. For two or more adjacent flipped plaquettes embedded in a positive background, only two zero modes persist, localized exclusively at the domain boundaries. As shown in Fig.~\ref{fig:mode}(b), a negative plaquette domain acts as a potential wall, mimicking open boundary conditions but defined at the level of plaquettes rather than sites. The corresponding energy spectrum [Fig.~\ref{fig:mode}(a)] reveals that in the ground‑state sector, additional flat bands emerge precisely at the positions where the all‑negative configuration produced flat bands.

\begin{figure*}
    \centering
    \includegraphics[width=\textwidth]{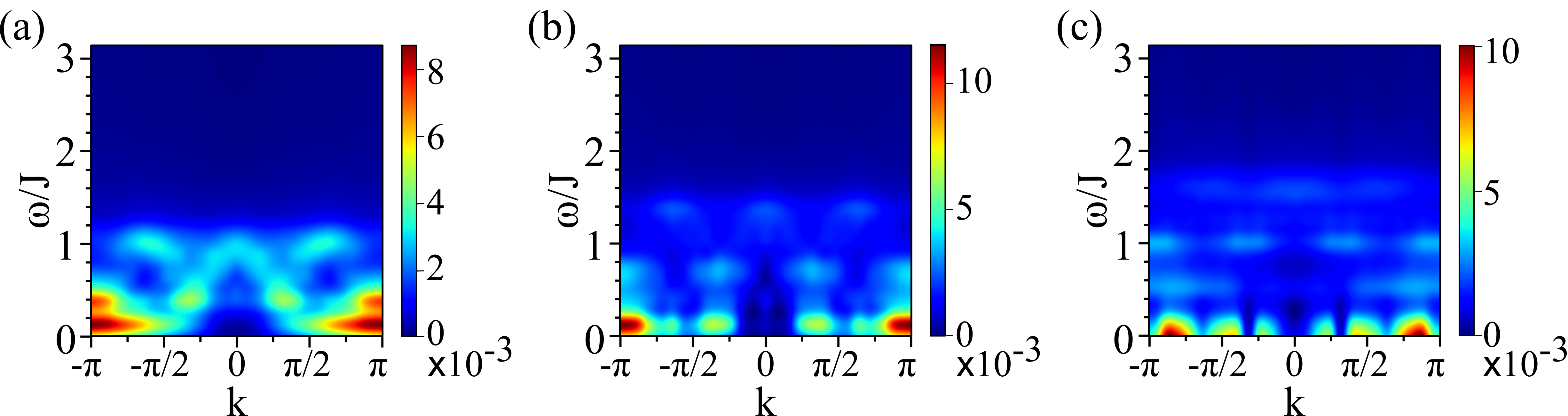}
    \caption{Single spin dynamical correlation function $(S^+(k,\omega))$: for (a) $J_r=1/\sqrt{2}$: in the trivial region, (b) $J_r=\sqrt{2}$: at the topological transition point and (c) $J_K=2$: inside the topological phase respectively. The results are obtained on a 24-site quasi-one-dimensional Kitaev chain using DMRG under OBC.}
    \label{fig:4}
\end{figure*}

A richer structure arises when multiple disconnected domains of negative plaquettes are introduced. Increasing the number of domains monotonically generates multiple pairs of near‑zero modes, with degeneracy growing alongside the number of negative plaquettes. For example, three domains of $-1$ plaquettes embedded in a positive background yield three pairs of near‑zero modes, each pair localized at the boundary of a single domain [Fig.~\ref{fig:mode}(c),(d)]. Extending this construction to a periodic arrangement of alternating two $-1$ and two $+1$ plaquettes across the chain produces a near-zero flat‑band spectrum, produced by $N/2$ pairs of states near the chemical potential.

These plaquette‑localized modes are robust throughout the trivial phase where the winding number vanishes. However, upon entering the topological regime, such modes disappear entirely: all plaquette sectors produce only extended bulk eigenstates. As illustrated in Fig.~\ref{fig:mode}(f), the energy spectrum at the gapless point (upper panel) and inside the topological region shows that dispersive bands vanish with fewer domain walls, and the gapless spectrum opens into a finite gap. This dichotomy highlights the interplay between plaquette excitations and global topology, with localized plaquette edge modes confined to the trivial phase and suppressed in the topological phase.

\subsection{Dynamical Spin Response and Fractionalized Excitations}

To characterize the low–energy excitations across different phases, we investigate the dynamical response generated by a single spin flip. As a probe, we compute the dynamical correlation function
\begin{equation}
    S^{+}(k,\omega)=-\frac{1}{\pi}\sum_{ij}\bra{\psi_{g}}S_{j}^{-}\frac{1}{\omega+i\eta-H}S_{i}^{+}\ket{\psi_{g}}\,e^{ik(r_{i}-r_{j})},
    \label{9}
\end{equation}
where $r_{i}$ denotes the site of the spin flip and $\eta$ is a small positive parameter ensuring causality. The correlation was evaluated using density matrix renormalization group (DMRG) simulations on a 24–site chain under open boundary conditions.

Fig.~\ref{fig:4} displays the spectral function for three representative couplings: $J_{r}=1/\sqrt{2}$, $J_{r}=\sqrt{2}$, and $J_{r}=2$. In all cases, the colormaps reveal broad continua rather than sharp dispersive magnon modes, indicating the fractionalization of spin excitations. This behavior resembles the broad continua observed in Kitaev systems and contrasts with conventional magnon excitations in ordered magnets. 

For $J_{r}=1/\sqrt{2}$ [Fig.~\ref{fig:4}(a)], the dominant spectral weight lies near momenta $k=\pm\pi$, slightly above zero frequency. At the critical point $J_{r}=\sqrt{2}$ [Fig.~\ref{fig:4}(b)], where the bulk gap closes, edge modes begin to emerge at zero frequency, but finite–size effects introduce a small residual gap. The spectral weight becomes more sharply localized at $k=\pm\pi$, yet the continuum character persists. Deep in the topological phase at $J_{r}=2$ [Fig.~\ref{fig:4}(c)], the response is essentially gapless, with the largest weight concentrated at $\omega\approx 0$. Unlike the critical case, however, the momentum distribution is spread across multiple values rather than confined to $\pi$, consistent with contributions from localized zero–energy edge states.

Taken together, these results demonstrate that the dynamical spin response across the transition is governed by fractionalized excitations. The persistence of continua, the redistribution of spectral weight, and the emergence of zero–frequency contributions in the topological regime provide evidence for Majorana‑like edge modes and a chiral topological phase transition in this quasi–one-dimensional chain.

\section{\label{sec:level4}Summary and Outlook}

In this work, we have introduced a quasi–one–dimensional spin chain with anisotropic Kitaev–like couplings, motivated by the plaquette structure of the original two–dimensional Kitaev model. Despite its reduced dimensionality, the chain retains a nontrivial plaquette geometry that admits a Majorana-fermion representation and renders the model integrable. As in the two–dimensional case, the emergent $\mathbb{Z}_{2}$ gauge fields determine the ground–state flux sector. The six–site unit cell hosts four non–dynamic Majorana modes, reflecting the underlying plaquette decomposition along the $x$–direction.

The Majorana band structure is fully gapped except at two critical values of the anisotropy parameter $J_{x}$: $J_{x}=0$ and $J_{x}=\sqrt{2J_{y}J_{z}}$, and general $J_{y}$ and $J_{z}$. The first corresponds to a trivial gapless point, while the latter marks a nontrivial gap closing associated with a chiral topological phase transition. For $J_{x}\geq\sqrt{2J_{y}J_{z}}$, the system enters a topological regime characterized by zero-frequency edge modes. Owing to its chiral block structure and the absence of additional interactions beyond Kitaev couplings, the chain falls within the BDI symmetry class of the Altland–Zirnbauer classification. This is corroborated by the quantized winding number and the numerical observation of edge states under open boundary conditions.

In the Lieb ground state plaquette sector, the model produces two flat and four dispersive bands, while the fully excited sector collapses into six flat Majorana bands. Introducing domains of negative plaquettes generates localized zero‑energy modes at domain boundaries, mimicking edge states in the trivial phase. Single or multiple flipped plaquettes yield robust boundary‑bound modes, with degeneracy increasing alongside the number of domains. These plaquette‑localized excitations persist only in the trivial regime and vanish in the topological phase where bulk states dominate. This highlights the interplay between plaquette domain walls and global topology.

To connect with potential material realizations, we note that compounds such as CuPzN, a quasi–one–dimensional Heisenberg chain with similar geometry, may provide a platform where anisotropic exchange interactions approximate the proposed Hamiltonian. Motivated by this, we investigated the dynamical spin correlation function to probe the low–energy excitations. DMRG simulations reveal broad continua in both the trivial and topological regimes, signifying fractionalization of spin excitations. In the topological phase, additional zero–frequency weight arises from edge modes, with spectral intensity spread across multiple momenta rather than confined to $k=\pi$. This contrasts with the sharp spinon continua of the Heisenberg chain and offers a distinctive fingerprint of the topological phase.

In summary, the quasi–one–dimensional Kitaev–like chain presented here exhibits a chiral topological phase transition characterized by anisotropy, fractionalized excitations across all regimes, and symmetry–protected edge modes in the topological phase. The coexistence of integrability, fractionalization, and topological protection highlights the model as a promising platform for exploring Majorana physics in reduced dimensions. Its dynamical response provides experimentally accessible markers that distinguish it from conventional Heisenberg spin chains, thereby suggesting possible routes towards identifying topological signatures in real materials.

\section*{\label{sec:level5}ACKNOWLEDGEMENTS}
 The author acknowledges financial support from
the Department of Atomic Energy (DAE), Govt. of India. RM also thanks I. Dutta, R. Parida, and S. Roy for valuable discussions, and also the VIRGO cluster, where the numerical calculations were performed.

\bibliography{manuscript}

\end{document}